# Etching of Photon Energy into Binding Energy in Depositing Carbon Films at Different Chamber Pressures

**Mubarak Ali**

Department of Physics, COMSATS University Islamabad, Islamabad Campus, 45550, Park Road, Pakistan, mubarak74@mail.com, mubarak60@hotmail.com, http://orcid.org/0000-0003-1612-6014

**Abstract:** A hot filament chemical vapor deposition is an attractive technique to deposit carbon films of different applications. In this technique, it is also feasible to study the influence of chamber pressure in the deposition of carbon films. In the deposition chamber, having dissociated from the methane precursor, gaseous carbon atoms first convert into the graphite state atoms and then into the diamond state atoms. An increase in the chamber pressure changes the morphology and structure of the deposited carbon films. The growth rate of the deposited carbon film increases by increasing the chamber pressure from 3.3 kPa to 8.6 kPa. The rate of converting gaseous carbon atoms into diamond atoms also increases. At 11.3 kPa and 14 kPa chamber pressure, gaseous carbon atoms convert into graphite state atoms at a high rate. The gas activation and gas collision processes vary broadly at varying chamber pressure. The morphology and structure of carbon films got deposited at different growth rates. The dissociation of molecular hydrogen into atomic hydrogen varies by varying the chamber pressure. Atomic hydrogen etched the photons released from the hot filaments. Thus, bits of differently shaped energy result. Gaseous carbon atoms convert into graphite and diamond state atoms under suitably shaped bits of energy. Graphite state atoms bind under the same involved bits, which is not the case when the diamond state atoms bind. The study sets a new trend in depositing, characterizing, and analyzing carbon films.

## 1.0. Introduction

Many research groups around the globe researched carbon films. The carbon films can deposit by employing various vapor deposition techniques. Some studies show the handling of deposition at low chamber pressures. Some studies show the selection of high chamber pressures. A hot filament chemical vapor deposition (HF-CVD) considers suitable to deposit carbon films. HF-CVD is simple for synthesizing carbon films.

There are many studies on carbon films, and some of them are cited here [1-8]. Carbon films with higher diamond content discussed some exciting aspects [9-14]. In the deposition of carbon films, switching dynamics under different conditions have been discussed [6]. The low growth rate of diamond film in HF-CVD has been the main drawback compared to the flame jet technique [15]. By employing the HF-CVD system, a carbon film was deposited by retaining with ~ 40 μm thickness [16].

A key parameter in the CVD diamond process is the system pressure, as it regulates the overall features of a carbon film. A systematic study on the influence of chamber pressure in depositing carbon films is yet not performed. Studies have reported that diamond film deposited at a low pressure contained more non-diamond components than those deposited at high pressure [17, 18]. Kang *et al.* [18] deposited well-faceted diamond cubes over silicon by keeping the chamber pressure at 0.13 kPa at the nucleation stage and chamber pressure raised to 2.7 kPa during the growth period. At different pressures, different diamond growth surfaces were studied [19].

Heimann *et al.* [20] studied the influence of different parameters on the deposition of carbon films. The growth rate of the carbon film at a chamber pressure of 200 kPa was about four times higher than the carbon film deposited at a chamber pressure of 50 kPa [21]. The set residence time also affects the growth rate [21, 22]. A faceted morphology developed at 1.3 kPa, and 2.7 kPa chamber pressures, whereas poor crystallinity at 0.7 kPa and 6.7 kPa chamber pressures [23]. The growth rate of diamond film increased between 1.2 kPa and 2.7 kPa [24]. Wan *et al.* [25] found that CVD diamond synthesis restricts by the gas phase composition, temperature, and pressure. Brunsteiner *et al.* [26] investigated the dependence of the growth rate on chamber pressure and found the maximum growth rate at chamber pressure was 2.7 kPa.

The deposition of carbon films can emerge with several applications. When the deposited carbon films are in the high-quality diamond phase, they can work efficiently for abrasives, cutting tools, and heat sink applications. When the deposited carbon films are rich in the graphite phase, they can work efficiently for electrode and field emission applications. When the deposited carbon films are equally good in diamond and graphite phases, they can work efficiently for the hybrid application.

The growth of carbon films, mostly diamond films, has been studied for decades. However, underlying science and engineering remained largely unaddressed. Again, there is incomplete information explaining the underlying science of carbon films. The discussion based on the science of influencing chamber pressures to deposit carbon films strikingly demands a comprehensive study. In the deposition of carbon films at different chamber pressures, the etching of photon energy into binding energy also discusses in preliminary detail.

## 2.0. Experimental Details

A Chinese-made HF-CVD system was employed to deposit the carbon films. The diameter of the wire is 0.5 mm, and the length of the wire is 13 cm. The distance between the substrate and filaments was set at ~ 7 mm. The approx. 10 mm distance of the parallel wire from the nearby parallel wire is maintained.

The base pressure was around $3\times10^{-3}$ Pa. A molecular turbopump maintained the base pressure.

The vane pump-2 also drained the gaseous species and maintained the deposition pressure for each set value.

This study studied the influence of varying chamber pressure from 3.3 kPa to 14 kPa. The total mass flow rate of the $H_2$ and $CH_4$ was 304.5 sccm during each experiment. Built-in mass flow controllers controlled the flow of $H_2$ and $CH_4$ gases. A methane concentration has 1.5% in each experiment. The approximate volume of the chamber was 2.9 cubic ft. Table 1 shows the detail of the chosen parameters in the deposition of different carbon films.

**Table 1:** Parameters of depositing carbon films at varying silicon substrates

| Chamber | Filament | Current (in amp)/ | Substrate | The flow rate of |
|---|---|---|---|---|

| pressure (kPa) | temperature (°C) | voltage (in volts) | Temperature (°C) | gases (sccm) |
| --- | --- | --- | --- | --- |
| 3.3 | 2100 | 255/12 | 895 | $CH_4$: 4.5, $H_2$: 300 |
| 6.0 | 2050 | 255/14 | 880 | $CH_4$: 4.5, $H_2$: 300 |
| 7.3 | 2030 | 255/16 | 905 | $CH_4$: 4.5, $H_2$: 300 |
| 8.6 | 2030 | 255/14 | 925 | $CH_4$: 4.5, $H_2$: 300 |
| 11.3 | 1800 | 255/12 | 820 | $CH_4$: 4.5, $H_2$: 300 |
| 14.0 | 1765 | 255/12 | 795 | $CH_4$: 4.5, $H_2$: 300 |

Polished p-type (100) silicon substrates (area 2×2 $cm^2$ and thickness 400 μm) were agitated ultrasonically. Substrates mechanically scratched for 5 minutes with suspension. The suspension prepared with diamond powder has a mesh size of ~ 30-40 (~ 28 μm) and acetone.

The silicon samples were scratched again for 10 minutes with the suspension prepared from acetone and diamond powder, a grain size of ~ 5 μm. The samples were washed with acetone to remove any debris from the substrate. In each new deposition, the installation of fresh filaments was made possible.

Input power of ~ 3.3 kW set to maintain the temperature of the filaments. There was no separate heater.

The temperature of the substrates is well-maintained by the heating effect of filaments. Temperature is approximately measured.

Both the optical pyrometer and K-type thermocouples were adjusted to measure the temperature. The rotational speed of the substrate holder was 5 rpm.

The processing time in each experiment was 10 hours. An initial 13 min spent to reach the set chamber pressure.

The X-ray reflectometer determined structural regularity in different carbon films (Philips PW3710). A copper line-focused X-ray tube having Kα radiation wavelength 1.541874 Å was used. The X-ray reflectometer is conventionally known as an X-ray diffractometer.

The field-emission scanning optical microscope used to analyze the surface and interface studies, FE-SOM (JEOL Model: JSM-7000F). The FE-SOM is conventionally known as FE-SEM.

In this work, the spectroscopic analyses of the carbon films performed by the micro-Raman spectroscope model HR800 UV have a He-Ne Red Laser wavelength of 632.8 nm.

## 3.0. Results and Discussion

### *3.1. Structural Examination of Carbon Films by X-ray Analysis*

Figure 1 (a-f) shows the X-ray patterns of the carbon films deposited at different chamber pressures. Carbon films show peaks at 2θ ~ 43.9°, ~ 75.3°, ~ 91.5°, and ~ 119.5°. Peaks are conventionally related to the (111), (220), (311), and (400) planes.

The peak at 2θ ~ 43.9° relates to the X-ray reflection from the outer ring electrons of the diamond atoms denoted by D (OR-R).

The peak at 2θ ~ 75.3° relates to the X-ray reflection from the zeroth ring electrons of the diamond atoms as denoted by D (ZR-R).

The peak at 2θ ~ 91.5° labeled by D (ZR-IR) relates to the inverse reflections of X-rays from the zeroth ring electrons of the diamond atoms. The peak at 2θ ~ 119.5° labeled by D (OR-IR) relates to the inverse reflections of X-rays from the outer ring electrons of the diamond atoms.

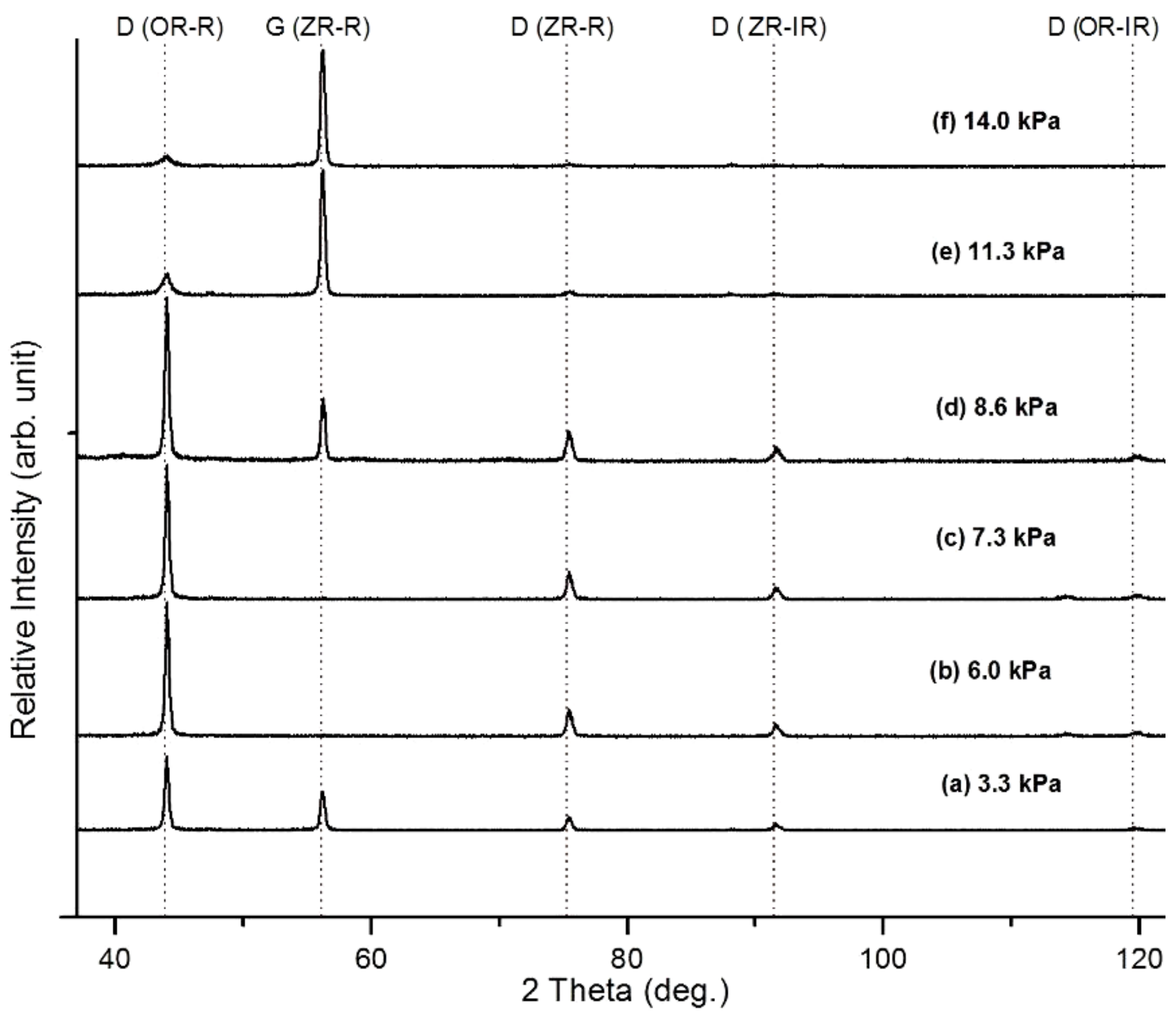

**Figure 1:** Structural determination of carbon films deposited at different chamber pressures, GIA = 0.5°

Labels (a) and (d-f) in Figure 1 also show a peak at 2θ ~ 56.1°. It is related to the graphite phase in deposited carbon films. The peak at 2θ ~ 56.1° has a very high intensity in carbon films deposited at chamber pressures of 11.3 kPa and 14.0 kPa, indicating the deposition of carbon films in the highly-graphitic phase. This peak also exists in the carbon film deposited at the lowest pressure. The peaks of the diamond phase almost diminish in the carbon films deposited at 11.3 kPa and 14.0 kPa. The peak at 2θ ~ 56.1° is related to the X-ray reflections from the zeroth ring electrons of the graphite atoms, as denoted by G (ZR-R) in Figure 1.

The peaks found in different patterns are related to the interaction of X-rays with atomic electrons of grains and particles belonging to the top-layered surface of carbon films [27]. The less-intensive graphitic peak appears in the patterns of carbon films deposited at the chamber pressures of 3.3 kPa and 8.6 kPa.

Carbon film deposited at 7.3 kPa chamber pressure keeps the highest structural regularity in terms of the diamond phase. The deposited carbon film does not show the peak of a graphite phase at 2θ ~ 56.1°.

However, reflected X-rays from the tiny grains, grains, and particles containing a diamond phase or a graphite phase print the associated peaks in the X-ray reflection (XRR) part and X-ray inverse reflection (XRIR) part of the X-ray pattern. Further detail on a structural determination by XRR analysis is given elsewhere [27].

### *3.2. Surface Morphology and Interface Study of the Carbon Films*

The carbon film grows uniformly at the lowest chamber pressure, which is 3.3 kPa. The deposited carbon film contains tiny grains, grains, and particles shown in Figure 2 (a). The morphology of the carbon film shows uniform distribution of tiny-sized grains, grains, and particles. Some grains and particles grow with the smooth faces in the carbon film. Many grains and particles also show their non-uniform growth behaviors. Due to the sharp edges of the crystallites, their growth shows different faces. In Figure 2 (A), many nucleated tiny grains do not grow further. Secondary nucleation starts.

At chamber pressure of 6.0 kPa, the dynamics of carbon atoms become favorable to grow tiny-sized grains and grains into the size of particles. Figure 2 (b) shows that the morphology of tiny grains, grains, and particles deviated toward the cubic growth behavior. In Figure 2 (B), a cubic growth behavior was more evident. Grains and particles were amalgamated in one region at a very high rate, as shown in Figure 2 (B). A particle is related to the size in the submicron range. A crystallite is related to shape.

At chamber pressure of 7.3 kPa, the growth behavior of grains and particles was in pyramid-shaped morphology. The growth behavior is uniform at a later stage, in Figure 2 (c). The pyramid-shaped grains and particles nucleated at the start of the process. It is evident in the interface image, as shown in Figure 2 (C).

At chamber pressure of 8.6 kPa, growth behavior is more dominant in the pyramid-shaped morphology of grains and particles. In Figure 2 (d), an image of the carbon film shows a tilted position. The morphology of grains and particles develops from the surface of the substrate, which could also be verified from the fractured cross-sectional view of the carbon film, as shown in Figure 2 (D).

At 11.3 kPa chamber pressure, grains and particles have dome-like shapes. It is evident in the fractured cross-sectional view of the carbon film shown in Figure 2 (E). The thickness of the carbon film synthesized at chamber pressure 11.3 kPa decreased several times compared to the carbon film deposited at chamber pressure 8.6 kPa. The thickness of the carbon film is even less than that of the carbon film deposited at a chamber pressure of 3.3 kPa. As labeled in the fractured cross-sectional view, the thickness is less than one micron, as shown in Figure 2 (E).

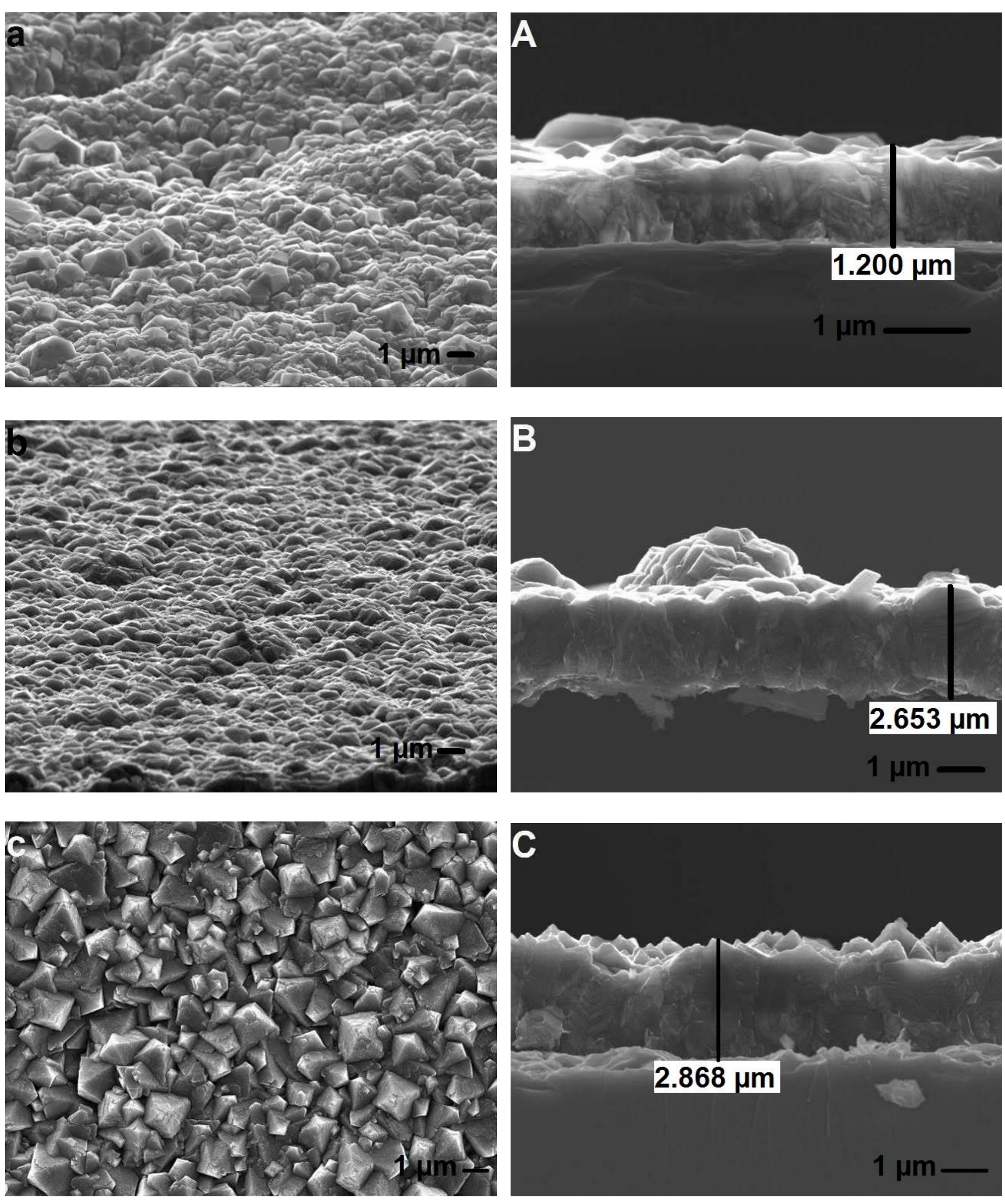

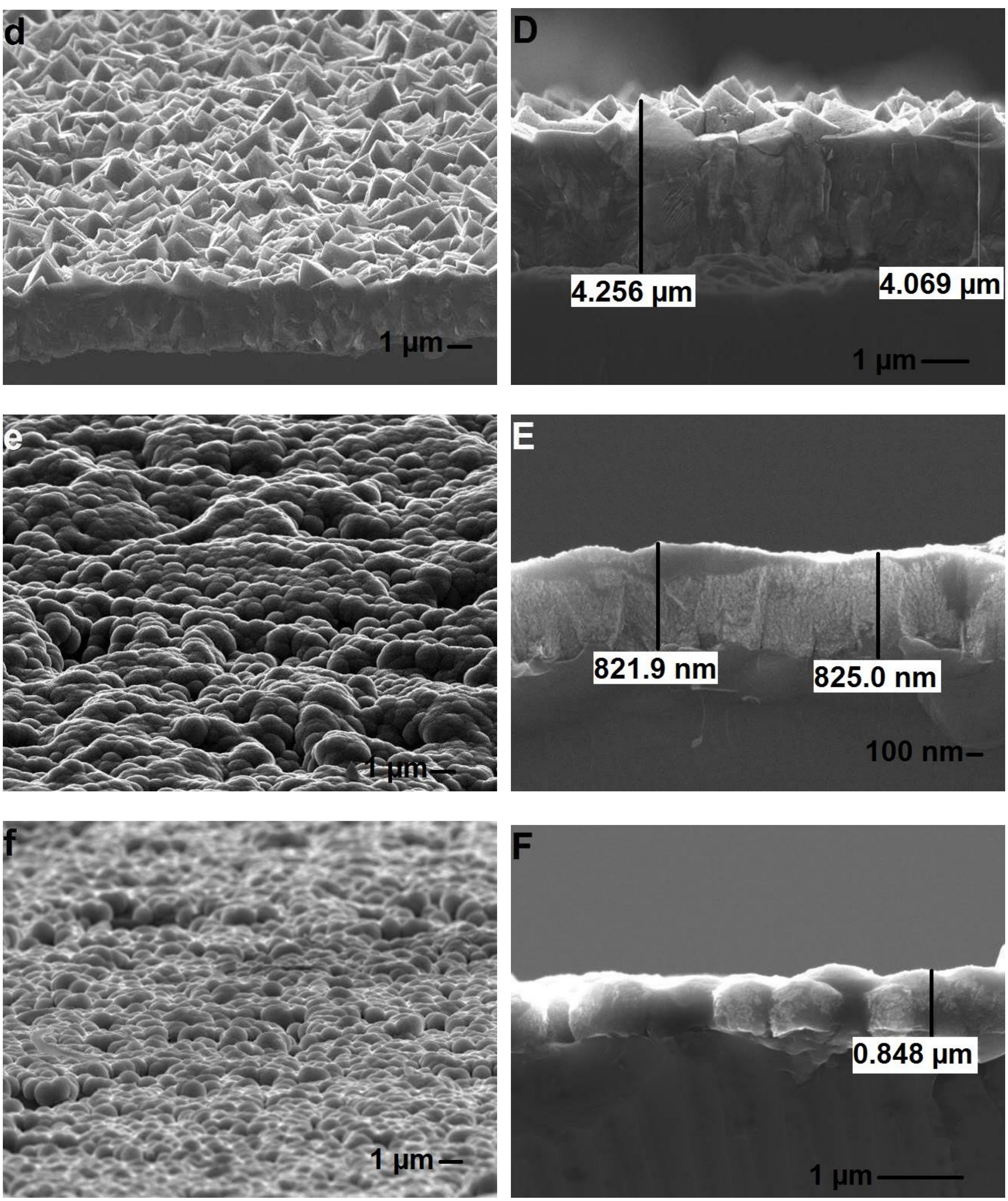

**Figure 2:** Surface morphology and fractured cross-sectional images of carbon films deposited at chamber pressure (a) 3.3 kPa, (b) 6.0 kPa, (c) 7.3 kPa, (d) 8.6 kPa, (e) 11.3 kPa and (f) 14.0 kPa

Grains and particles again show dome-like shapes at 14.0 kPa chamber pressure. The thickness of the carbon film deposited at chamber pressure 14.0 kPa is the same as in the case of a carbon film deposited at chamber pressure 11.3 kPa. The labeled thicknesses keep slightly different values in the fractured cross-sectional views of the carbon films shown in Figures 2 (E) and 2 (F).

By increasing the chamber pressure from 5.3 kPa to 39.5 kPa, the morphology of the carbon film varied [20]. However, the primary mechanism of the change of morphology remained unclear. Schwarz *et al.* [22] observed the pressure-dependent growth rate of diamond coatings at a substrate temperature of 850 °C. On the other hand, the growth rate was 0.2 μm/h at a chamber pressure of 5.0 kPa. At 0.7 kPa chamber pressure, a bit lower growth rate was noted [26].

At low chamber pressures, our results agree reasonably with the chamber pressures 2.0 kPa to 5.0 kPa as given elsewhere [22]. At the highest value of the chamber pressure, the growth rate decreased below 0.05 μm/h [26]. However, in the present study, the highest growth rate (~ 0.4 μm/h) obtains at a chamber pressure of 8.6 kPa, which drastically decreases at chamber pressures of 11.3 kPa and 14.0 kPa (~ 0.1 μm/h). The concentration of atomic hydrogen near the filament surface has been discussed earlier [22, 28]. The results of Brunsteiner *et al.* [26] contradict the growth rates of carbon films found in the results of Schwarz *et al.* [22]. The growth rate and quality of the diamond at atmospheric pressure (1059.1 kPa) were better than the chamber pressure of 5.0 kPa [29].

### ***3.3. Raman Spectroscopic Analyses of the Carbon Films***

In different Raman spectra of the deposited carbon films, the recorded intensity of the Raman signals reveals the morphology and structure of tiny grains, grains, and particles. The presence of the Raman peak near wave number 1332.1 $cm^{-1}$ corresponds to the diamond peak, as shown in Figure 3. In Figure 3, the prominent peak shows different trends and features of the deposited carbon films due to varying chamber pressure.

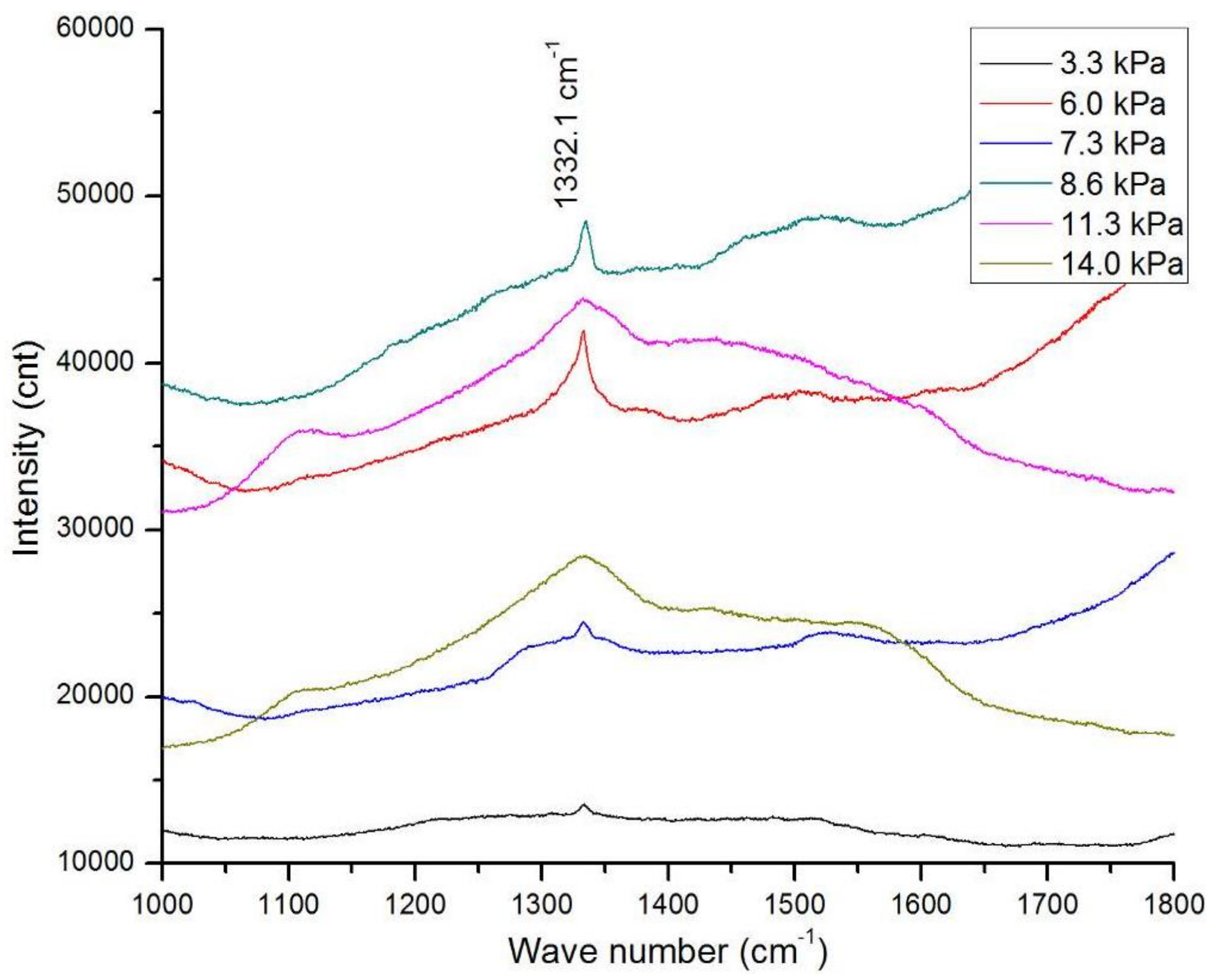

**Figure 3:** Raman spectra of carbon films deposited at different chamber pressures

In Figure 3, a slight shift in the peak is due to the associated stress of the carbon film. The Raman line width varied from the mode of the deposited diamond [30]. At wavenumber 1581 $cm^{-1}$, an additional peak originated in Figure 3. It is related to the G-peak declared due to the highly orientated graphitic phase in CVD diamond coatings [31]. A review discussed the structure of carbon films by Raman spectroscopy [32].

The carbon films synthesized at different chamber pressures are sensitive to the energy signals shown in Figure 3. A pronounced peak at wave number 1100 $cm^{-1}$ is found in the carbon films deposited at chamber pressures of 11.3 kPa and 14.0 kPa, indicating the presence of a high graphitic phase. It is a need to re-investigate the Raman spectroscopy of the carbon films.

### *3.4. Some Aspects of Nucleation and Growth Mechanisms of Carbon Films*

Graphite atoms might not adhere to the seeded species. Thus, when the carbon atoms deposit in the graphite state, they can detach from the substrate surface. Graphite atoms might not trap in the variable regions of the substrate. There is a probability of detachment of those graphite state atoms from the substrate, so nucleation of tiny-sized grain either in graphite state atoms or in diamond state atoms starts from good sites of the seeded and scratched substrate. It indicates that depositing carbon atoms in different states by fully covering the substrate is a grand challenge.

The mean free path of active species is inversely proportional to the pressure as per the kinetic theory of gases.

The mean free path of active species decreases by increasing the chamber pressure. At the start of the process, fluctuation of the parameters takes place in each process. It is mainly due to the access to set conditions of the process. Again, fluctuation in parameters can be due to the water circulation. It can be due to the change in the positions of filaments. They can influence the initial nucleating layer of the carbon film. So, a different time duration to achieve set chamber pressure can cause a different nucleation rate.

Lee *et al.* [17] predicted the concentration of active species first increases exponentially and decreases by decreasing the pressure finally. The carbon atoms, which survive, reach the substrate surface to deposit. So, carbon atoms, which do not position to deposit at the substrate surface, either become part of the drain or contaminate the process of nucleating tiny-sized grains. The nucleation of tiny-sized grains also depends on the rate of releasing bits of energy from the gas collision. Keeping the same nucleation rate in depositing carbon films is a great challenge.

An increase in the mean free path leads to partial control of the dynamics in the depositing carbon atoms. It is mainly in the case of low chamber pressures. Therefore, a deposited carbon film keeps the growth partially diamond-specific and partially graphite-specific.

The temperature of the filaments decreases at high chamber pressure, as shown in Table 1, under fixed input power. As there is no separate heater to heat the substrate, the substrate temperature also decreases. It is a cause of enhancement of the secondary nucleation. Thus, the thicknesses of the deposited carbon films decreased, as shown in Figures 2 (E) and (F). Figure 2 (D) shows that the carbon film thickness deposited at a chamber pressure of 8.6 kPa remains the highest. The conversion rate of molecular hydrogen into atomic hydrogen increases at 8.6 kPa chamber pressure. More gaseous carbon atoms convert into a diamond state. However, the conversion rate of molecular hydrogen into atomic hydrogen decreases at 11.3 kPa and 14.0 kPa chamber pressures. So, carbon atoms convert into diamond state atoms at a low rate. The growth rates of carbon films decrease along with the deteriorating quality of the diamond content.

Under the commonly employed conditions of the HF-CVD process, the transportation of atomic hydrogen for the growing surface is a diffusion-limited process [33]. Diamond grown by HF-CVD investigated the growth mechanism from the surface topography [34].

The growth of diamond is a sliding scale between atomic hydrogen and hydrocarbon radical [35]. Substrate temperature and pressure jointly influence the growth rate of diamond films [36].

At high chamber pressures, the dissociation of methane into carbon atoms is high compared to the low and intermediate chamber pressures. However, dissociated carbon atoms frustrate as a sufficient amount of carbon atoms remain in the gaseous state. So, the depositing rate of carbon atoms in the graphite state atoms becomes low. Consequently, the rate of converting carbon atoms into diamonds becomes low.

### *3.5. Growth Rate versus Chamber Pressure*

Figure 4 shows the growth rates of diamond and graphite phases in deposited carbon films at different chamber pressures. The estimated zone of the depositing carbon film at a nearly equal growth rate of diamond and graphite is also labeled by (1) in Figure 4.

The maximum diamond growth rate obtained at 8.6 kPa chamber pressure is labeled by (2) in Figure 4.

The maximum graphite growth rate obtained at 14.0 kPa chamber pressure is labeled by (3) in Figure 4.

The dynamics of carbon atoms work in different manners at 3.3 kPa chamber pressure. Many gaseous carbon atoms get converted into diamond state atoms at intermediate chamber pressures, 7.3 kPa and 8.6 kPa. At high chamber pressures, the morphology of grains and particles develops in the dome shape. However, the growth rates of carbon films are low. The gaseous carbon atoms mainly get converted into graphite content. At 11.3 kPa, the morphology of grains is more like the dome shape. On increasing the chamber pressure up to 14.0 kPa, the morphology of grains is again like the dome shape.

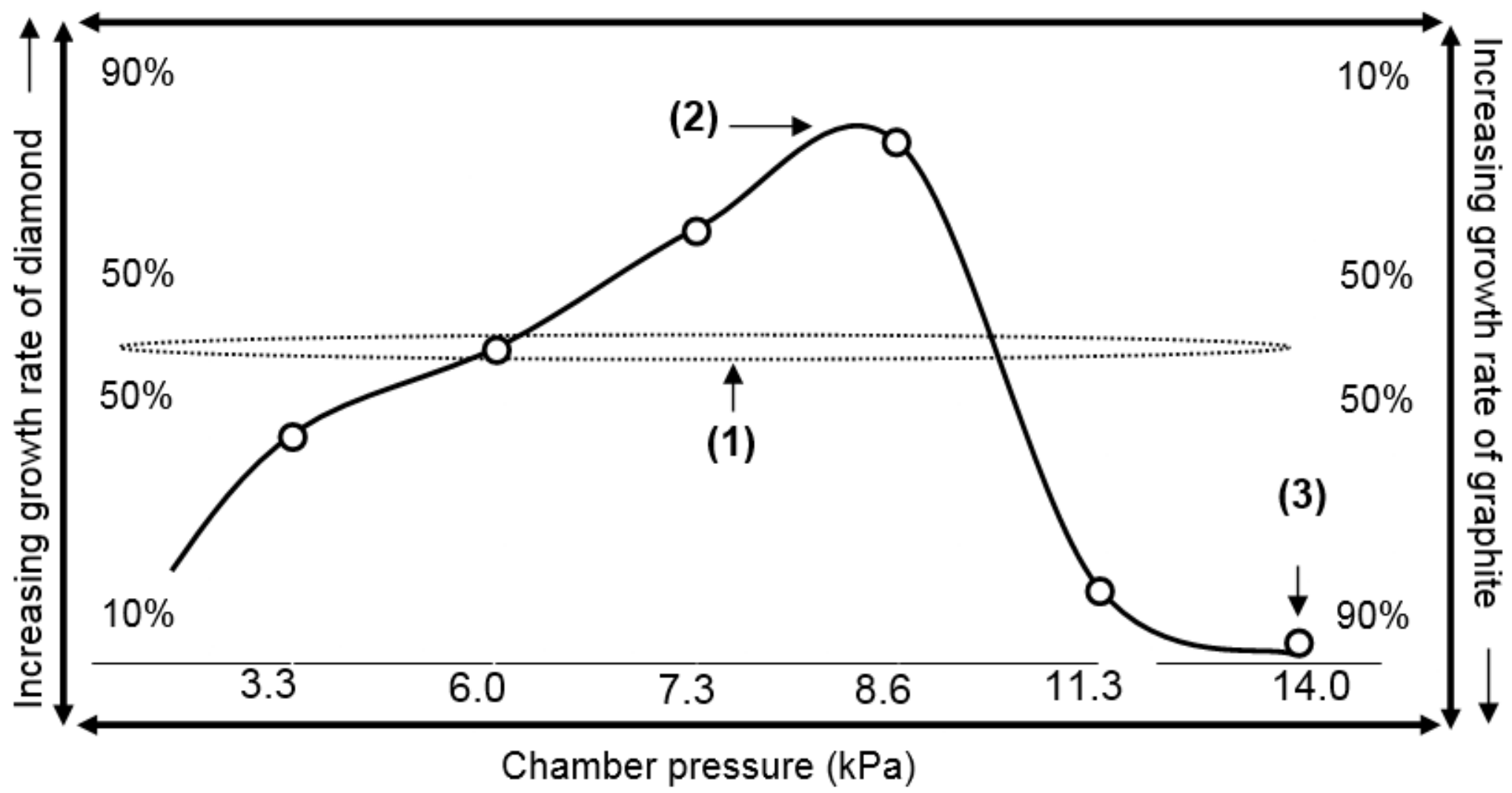

**Figure 4:** Growth rates of diamond and graphite in the carbon films deposited at varying chamber pressure; **(1)** a zone of balanced (or equal) growth rates of diamond and graphite, **(2)** a zone of the high growth rate of diamond and **(3)** a zone of the high growth rate of graphite

At high chamber pressures, locally engaged forces synchronize to preserve the bits of shaped energy required to form tiny grains, grains, and particles with a graphitic structure. However, locally engaged forces synchronize at intermediate chamber pressures to preserve the bits of shaped energy required to form tiny grains, grains, and particles with a diamond structure.

### *3.6. Dissociation of $H_2$ and $CH_4$ Gases – A Gas Activation Process*

The dissociation rates of $H_2$ and $CH_4$ gases determine the gas activation process. In the HF-CVD reactor, molecules of methane and hydrogen dissociate for conversation into atomic states. Molecular hydrogen converts into hydrogen atoms under the thermal activation of hot filaments. In converting the methane molecule into the carbon component and the four hydrogen atoms, the hydrogen atoms detach from the carbon atom. On detachment of the four hydrogen atoms, the binding states of a carbon atom become unfilled, enabling its conversion to either graphite or diamond state atom.

In the dissociation of the $CH_4$ molecule, expansion and contraction behaviors of two different natures are triggered, where the carbon component gives away four hydrogen atoms. $CH_4$ converts into a carbon component and four hydrogen atoms. In the process of dissociation of the $CH_4$ molecule, four hydrogen atoms separate at the same time. However, a hydrogen molecule requires activation energy up to a higher level to convert into hydrogen atoms. A study on diamond films discusses the temperature of hot filaments in the deposition [37].

The dissociation rate of molecular hydrogen is varied at different chamber pressures. Again, the conversion rate of methane molecules into gaseous carbon and hydrogen atoms becomes different under the varying chamber pressure. The gas activation process varies at each chamber pressure. The rate of etching photon energy also varies at each chamber pressure.

There is a need to study the dissociation rate of $H_2$ and $CH_4$ gases at varying chamber pressure. Again, to find out the different species of gas activation processes at different chamber pressures, there is also a need to study the optical emission spectra.

Some other qualitative and quantitative analyses can be employed to investigate the rate of gas dissociation and the gas activation process.

### *3.7. Etching of Photon Energy into Bits of Dash and Golf-Stick-Shaped Energy*

A photon of short length called an overt photon is shown in Figure 5 (a). Atomic hydrogen etches the overt photon when traveling oppositely to it. A description mechanism of the hydrogen atom under the new insight is given elsewhere [38].

To etch a short-length photon, both electronic tips of the atomic hydrogen involves. As a result, that overt photon converts into pieces or bits of dash-shaped energy. Figure 5 (a) shows it.

Figure 5 (b) shows a photon etching by atomic hydrogen. In Figure 5 (b), parabola-shaped energy bits form. Atomic hydrogen utilizes one electronic tip to develop parabola-shaped energy bits. Figure 5 (b) shows it.

Atomic hydrogen also utilizes one electronic tip to develop golf-stick-shaped energy bits. The bits of parabola-shaped energy convert into bits of golf-stick-shaped energy. Figure 5 (c) shows the etching of parabola-shaped energy bits.

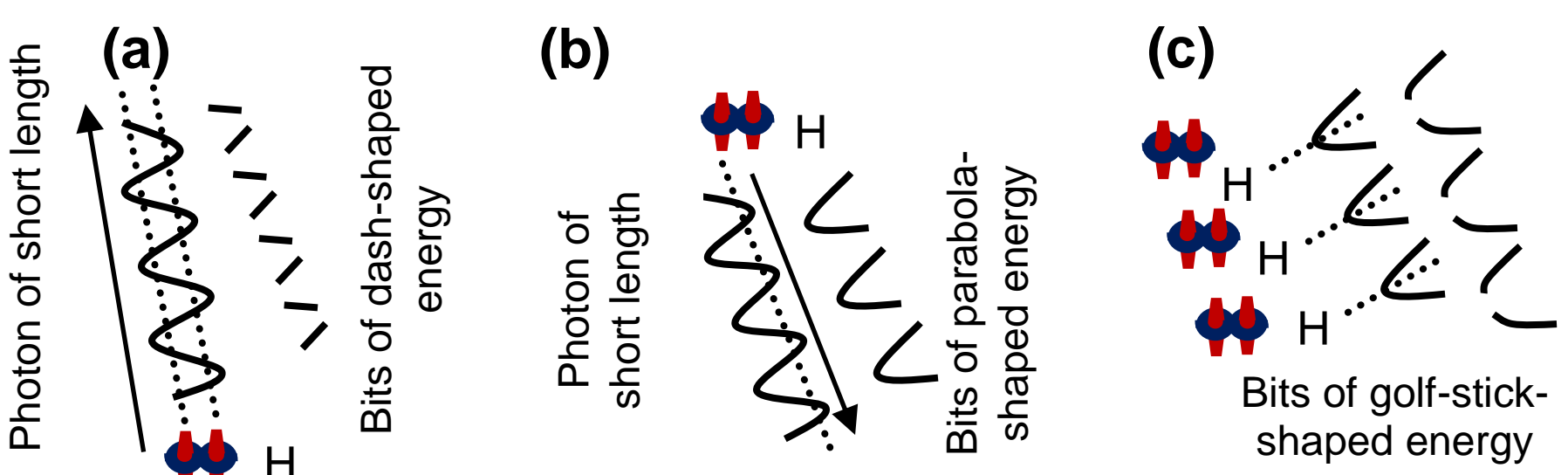

**Figure 5:** (a) etching of a short-length photon into bits of dash-shaped energy, (b) etching of a short-length photon into bits of parabola-shaped energy, and (c) etching of the parabola-shaped bits into golf-stick-shaped energy bits

Dash-shaped bits convert gaseous carbon atoms into graphite and diamond state atoms. Graphite state atoms bind under the previously involved dash-shaped energy. However, this is not the case for diamond atoms. A golf stick-shaped energy is involved in the binding of diamond state atoms.

Atomic hydrogen etched the photon energy rather than etched the carbon atoms. Therefore, it is not correct that atomic hydrogen etched carbon. It is atomic hydrogen, which etches photon energy. Figure 5 (a) shows the etching of photon energy into dash-shaped energy. Figure 5 (b) shows the etching of photon energy into parabola-shaped energy.

Dash-shaped energy of two bits involves transferring the electrons converting the carbon atom into another state [39]. Gaseous carbon atoms get converted into graphite and diamond state atoms in the region close to the hot filaments. The distance between the substrate and hot filaments is a few millimeters. The suitable interaction of atomic hydrogen can further etch the bits of parabola-shaped energy. The shape of a new bit of energy becomes like a golf stick. In the diamond, atoms bind by the involvement of golf-stick-shaped energy bits [39].

A filament becomes hot due to the propagation of a photonic current. In the deposition chamber, hot-filaments release photons of short and long lengths called overt photons. Under the etching of atomic hydrogen, a photon is etched into the bits of energy, having the shape of a dash, parabola, and golf stick. A separate study has detailed the photon study [40].

### *3.8. Gas Collision and Releasing Different-Shaped Bits of Energy*

The etching of photon energy for bits of golf-stick-shaped energy increases by increasing the chamber pressure up to 8.6 kPa. It decreases by increasing the chamber pressure up to 14.0 kPa, shown in Figure 6 (a) by the solid curve. The etching of photon energy into dash-shaped energy bits remains moderate, low, and high at the lowest, intermediate, and high chamber pressure. In the dotted curve, Figure 6 (b) shows the etching of photon energy into dash-shaped energy at different chamber pressures.

In label (1) of Figure 6 (a), the blue dotted line shows the point of gas collision where the minimum disruption in the formation of the parabola and golf-stick-shaped energy bits takes place. Therefore, gas collision is more favorable at intermediate chamber pressures to release bits of golf-stick-shaped energy. Hence, carbon films have a higher growth rate than the diamond phase.

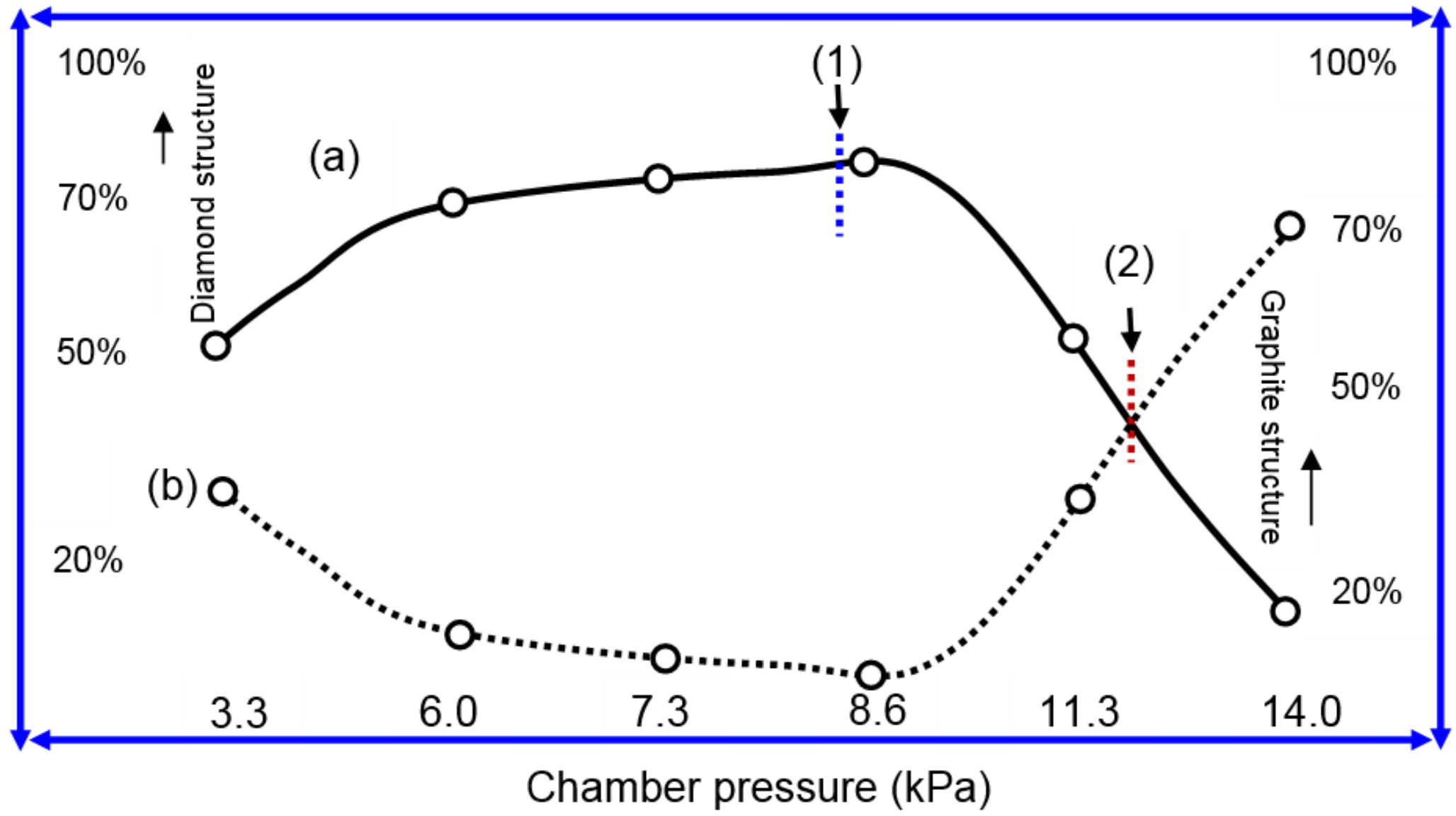

**Figure 6:** (a) pressure-dependent growth showing the point of gas collision having the minimum disruption in coming golf-stick-shaped energy to bind (1) diamond atoms. (b) Pressure-dependent growth showing the point of a gas collision having the minimum disruption in dash-shaped energy coming to bind atoms of (2) graphite

In a gas collision, atomic hydrogen and gaseous carbon are involved, but molecular hydrogen and methane are also involved. The collision rate of gases varies largely at varying chamber pressure, as shown in Figure 6 in estimation. However, the collision rate of different gases should first depend on the process of gas activation.

In label (2) of Figure 6 (b), the red dotted line shows the point of gas collision where the minimum disruption in the formation of bits of dash-shaped energy takes place. The gas collision is favorable to releasing the bits of dash-shaped energy.

The deposition chamber has less photon energy at 11.3 kPa and 14.0 kPa chamber pressures. It is due to the excessive contamination of filaments. So, the conversion of atomic hydrogen remained at a lower rate. Thus, the rate of converting gaseous carbon atoms into diamond atoms remained minimum. Again, high chamber pressures deal with the more dissipation of heat energy instead of dealing with the more releasing of photon energy. The presence of heat energy inside the reactor affects the process.

### *3.9. General Discussion*

In studies given elsewhere [2, 9], the role of atomic hydrogen remained crucial in determining the content-specific growth of a carbon film. Increasing a small amount of $CH_4$ in the total mass flow rate lowers the quality of diamond film [2]. These studies validate the crucial role of gas collision.

In Figures 2 (c) and 2 (d), the tilted microscopic images show the pyramid-shaped morphology of large diamond crystallites. Many diamond crystallites are grown by retaining their roots in contact with the substrate surface. In the growth of each diamond crystal, a large number of gaseous carbon atoms got converted into diamond state atoms. Therefore, the carbon films relate to more diamond phases. Figures 2 (c) and (d) show an almost similar trend of growth behavior of grains and particles. Such kind of growth behavior discloses the stinging nature of diamonds. Diamond growth is south to ground, but the structure of diamond is tetra-electron ground to south topological structure [39].

Under different arrangements of hot filaments, the growth rate of the carbon film may differ from the one discussed here. In microwave-based chemical vapor deposition, the structure of tiny grains of carbon films explained the nucleation behavior with some essential detail and how secondary nucleation was initiated [41]. Therefore, much more work is required in the HF-CVD technique to understand the nucleation and growth mechanisms.

Many research groups investigated the applications of deposited carbon films [42-45]. The ratio of $sp^2$ to $sp^3$ is low at the intermediate set chamber pressures, whereas it is high at the high set chamber pressures, which is evident in Figure 4. In the current study, $sp^2$ relates to the graphitic phase, whereas $sp^3$ relates to the diamond phase. The ratio of $sp^2$ to $sp^3$ in carbon films has been studied earlier [46, 47].

To keep constant filament and substrate temperatures is quite a difficult task. It happened due to varying chamber pressure in the reactor. The published studies also indicate the approximate values of temperature. A different design of the chamber affects the pressure differently. There is a need to discuss the kinetics, dynamics, mass spectroscopy, computational modeling of gas dissociation, and the gas activation of the process. Atomic hydrogen production on the filament and its diffusion to the substrate surface need to explore further. The chemical kinetics of the process at different steps also need to explore further.

## 4.0. Conclusion

Tiny carbon grains nucleate at different rates for different chamber pressures. At low chamber pressures, the rate of converting gaseous carbon atoms into graphite and diamond state atoms is moderate. At intermediate chamber pressures, gaseous carbon atoms convert into diamond atoms at a high rate. The conversion rate of gaseous carbon atoms for graphite state atoms increases at high chamber pressures. Nucleations of the diamonds start at the feasible sites of the roughened substrate. In the carbon film, tiny grains grow, reaching the size of grains and particles.

The morphologies of grains and particles at low, intermediate, and high chamber pressures develop in the mixed, pyramid, and dome shapes, respectively. The thickness of the carbon films depends on the rate of converting gaseous carbon atoms into different states. The morphologies of the grains and particles are sensitive to the Raman signals giving information about the shapes. Grains and particles show high purity diamond-wise at chamber pressures 6.0 kPa, 7.3 kPa, and 8.6 kPa, whereas high purity graphite-wise at chamber pressures 11.3 kPa and 14.0 kPa. X-ray analyses of the carbon films deposited at varying chamber pressures agree with the information provided by the spectroscopic analysis and FE-SOM investigation.

Dissociating $CH_4$ molecules and molecular hydrogen varies at different chamber pressures. A $CH_4$ molecule simultaneously dissociates into a carbon atom and four hydrogen atoms, and a hydrogen molecule dissociates into the atomic hydrogen under the thermal activation of the hot filaments. The rate of dissociation of gases determines the gas activation process. The rate of the gas activation process determines the gas collision process.

The hot filaments release the photons at different rates at different chamber pressures. Due to the varying gas activation and gas collision processes at each chamber pressure, the amount of photon energy etched into bits of the dash and golf-stick-shaped energy is also varied. The conversion rate of the gaseous carbon atoms into graphite and diamond state atoms depends on the formation of atomic hydrogen.

The gas activation and gas collision process at different chamber pressures influence the binding rate of carbon atoms. At intermediate chamber pressures, more molecular hydrogen dissociates into atomic hydrogen. Thus, the photon energy converts into bits of golf-stick-shaped energy in higher amounts. At chamber pressures 6.0 kPa, 7.3 kPa, and 8.6 kPa, the gaseous carbon atoms convert into diamond state atoms at a high rate, but the rate of their binding also becomes high.

Bits of dash-shaped energy convert gaseous carbon atoms into another state, the graphite state atoms bind under the same bits of dash-shaped energy, and the diamond state atoms bind under the bits of golf-stick-shaped energy [39]. At intermediate chamber pressures, a gas collision process favors the etching of the photon energy. So, the bits of golf-stick-shaped energy produce in higher amount. The etching of photon energy in the dash shape is high at 11.3 kPa and 14.0 kPa. At 6.0 kPa, 7.3 kPa, and 8.6 kPa, the etching of photon energy in parabola and golf-stick shapes is high. At 3.3 kPa, the etching photon energy in parabola and golf-stick-shaped energy is moderate.

The work provides insight into problem-based solutions. It can enable one to explore cutting-edge technologies. There is vast room to synthesize carbon films with different applications in the hot-filaments reactor. New designs of the deposition systems can provide further opportunities to synthesize carbon films. In new depositions, resource utilization can minimize, leading to the building of a clean environment.

**Acknowledgments:**

Mubarak Ali expresses sincere gratitude to The Scientific and Technological Research Council of Türkiye (TÜBİTAK), the letter ref. # B.02.1.TBT.0.06.01-216.01-677-6045 for honoring the postdoc award for the year 2010. Mubarak Ali thanks ex-postdoc advisor Professor Dr. Mustafa Ürgen (now retried from Istanbul Technical University) for continuous research support. He also thanks Mr. Talat ALPAK (ITU, Istanbul) for helping in the field emission scanning microscope. Mubarak Ali appreciates the kind support of Professor Dr. Kürşat KAZMANLI, Dr. Erdem Arpat, Dr. Semih ÖNCEL, Dr. Manawer, Dr. Nagihan Sezgin, Dr. Sinan Akkaya, and others while staying at ITU.

**Data Availability Statement:**

All data generated or analyzed during this study are included in this published article.

**Conflicts of Interest:**

The authors declare no conflicts of interest.

## Author's biography:

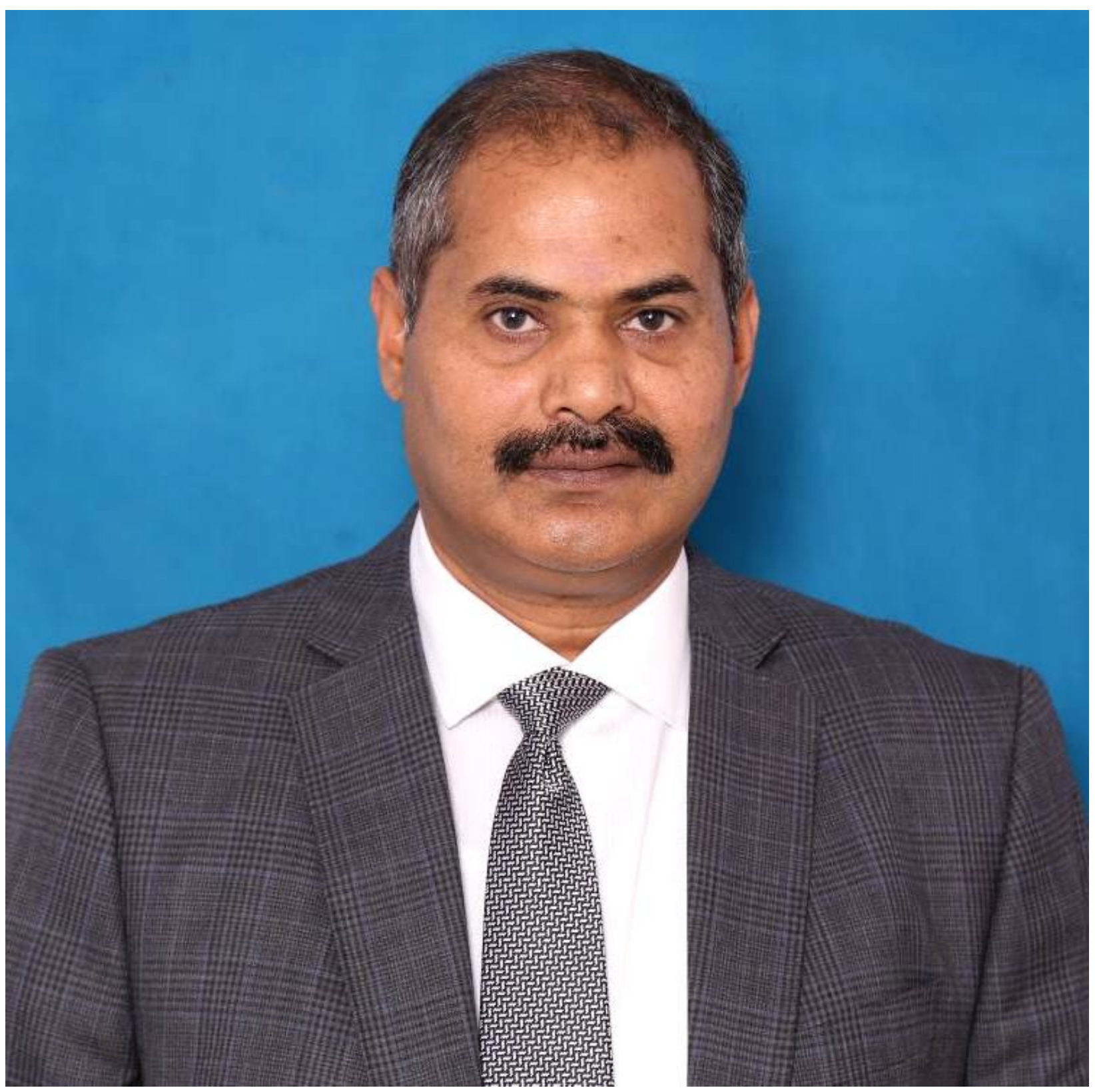

**Mubarak Ali** graduated from the "University of the Punjab" with BSc (Phys & Math) in 1996 and MSc in Materials Science with distinction from Bahauddin Zakariya University, Multan, Pakistan (1998); his thesis work was completed at Quaid-i-Azam University Islamabad. He gained a Ph.D. in Mechanical Engineering from the Universiti Teknologi Malaysia under the award of the Malaysian Technical Cooperation Programme (MTCP;2004-07) and a postdoc in advanced surface technologies at Istanbul Technical University under the foreign fellowship of The Scientific and Technological Research Council of Turkey (TÜBİTAK, 2010). Dr. Mubarak completed another postdoc in nanotechnology at the Tamkang University Taipei in 2013-2014, sponsored by National Science Council, now M/o Science and Technology, Taiwan (R.O.C.). Presently, he is working as an Assistant Professor on the tenure track at COMSATS University Islamabad (previously known as COMSATS Institute of Information Technology), Islamabad, Pakistan (since May 2008, and the position renewal is in the process) and before that, worked as assistant director/deputy director at M/o Science & Technology (Pakistan Council of Renewable Energy Technologies, Islamabad, 2000-2008). The Institute for Materials Research, Tohoku University, Japan, invited him to deliver a scientific talk. He gave several scientific talks in various countries. His core area of research includes materials science, physics & nanotechnology. He was also offered a merit scholarship for Ph.D. study by the Higher Education Commission, Government of Pakistan, but he did not avail himself of the opportunity. He earned a diploma (in English) and a certificate (in the Japanese language) in 2000 and 2001, respectively, part-time from the National University of Modern Languages, Islamabad. He is the author of several articles available at the following links;
https://www.researchgate.net/profile/Mubarak_Ali5 &
https://scholar.google.com.pk/citations?hl=en&user=UYjvhDwAAAAJ